\documentclass[cits]{PoS}
\newcommand{\SU}{\mathrm{SU}}
\newcommand{\Tc}{T_{\mbox{\tiny{c}}}}
\newcommand{\SW}{S_{\mbox{\tiny{W}}}}
\newcommand{\nconf}{n_{\mbox{\tiny{conf}}}}
\newcommand{\Tr}{{\rm Tr\,}}

\title{Conformal perturbation description of deconfinement}
\ShortTitle{Conformal perturbation}

\author{\speaker{Michele Caselle}\\
        Department of Physics and Arnold-Regge Center, University of Turin, and INFN, Turin\\
        Via Pietro Giuria 1, I-10125 Turin, Italy\\
        E-mail: \email{michele.caselle@to.infn.it}}

\author{Nicodemo~Magnoli\\
        Department of Physics, University of Genoa, and INFN, Genoa\\
        Via Dodecaneso 33, I-16146, Genoa, Italy\\
        E-mail: \email{nicodemo.magnoli@ge.infn.it}}

\author{Alessandro~Nada\\
        NIC, DESY\\
        Platanenallee 6, D-15738 Zeuthen, Germany\\
        E-mail: \email{alessandro.nada@desy.de}}

\author{Marco~Panero\\
        Department of Physics, University of Turin, and INFN, Turin\\
        Via Pietro Giuria 1, I-10125 Turin, Italy\\
        E-mail: \email{marco.panero@unito.it}}

\author{Marcello~Scanavino\\
        Department of Physics, University of Genoa, and INFN, Genoa\\
        Via Dodecaneso 33, I-16146, Genoa, Italy\\
        E-mail: \email{marcello.scanavino@ge.infn.it}}
\abstract{
Conformal perturbation theory is a powerful tool to describe the behavior of statistical-mechanics models and quantum field theories in the vicinity of a critical point. In the past few years, it has been extensively used to describe two-dimensional models and recently has also been extended to three-dimensional models. We show here that it can also be used to describe the behavior of four-dimensional lattice gauge theories in the vicinity of a critical point. As an example, we discuss the two-point correlator of Polyakov loops close to the thermal deconfinement transition of $\SU(2)$ Yang-Mills theory. We show that the short-distance behavior of this correlation function (and, thus, of the interquark potential) is  described very well by conformal perturbation theory. This method is expected to work with a similarly high accuracy for all critical points in the same universality class, including, in particular, the critical endpoint in the QCD phase diagram.}

\FullConference{XIII Quark Confinement and the Hadron Spectrum - Confinement2018\\
		31 July - 6 August 2018\\
		Maynooth University, Ireland}

\begin{document}

\section{Introduction}

In order to understand the properties of confinement, a precise quantitative characterization of the interquark potential in non-Abelian gauge theories is of crucial importance. From an analytical point of view, this is a non-trivial task, which requires a careful combination of perturbative analysis at short interquark distances $r$, and some effective description (e.g. in terms of a string model~\cite{Kuti:2005xg,Lucini:2012gg,Brandt:2016xsp}) at large distances. High-precision numerical estimates of the interquark potential can be obtained in a straightforward way from Monte~Carlo calculations on the lattice, and are very useful to test our understanding of the ultraviolet and infrared physics of these theories. In the limit in which the temperature $T$ is close to zero, the situation is well understood. The short-distance behavior is described well by standard perturbation theory, while the large-distance behavior can be accurately modeled by a confining bosonic string. In particular, the latter includes universal corrections proportional to $1/r$ (the well-known L\"uscher term~\cite{Luscher:1980ac,Luscher:1980fr}) and to $1/r^3$~\cite{Luscher:2004ib}, which can also be accurately unambiguously identified in lattice simulations~\cite{Caselle:2016wsw}. The combination of perturbative calculations at short distances and effective-string modeling at large distances yields a very accurate description of the interquark potential in the low-temperature regime of Yang-Mills theories~\cite{Necco:2001xg,Lohmayer:2012ue,Husung:2017qjz}.

The picture, however, is less clear at finite temperature, especially when $T$ is close to the deconfinement temperature $\Tc$. In particular, in this regime the long-distance behavior of the potential features non-universal, higher-order (in powers of $1/r$) terms of the effective-string model. On the other hand, this regime is particularly interesting from a physical point of view, due to the large amount of experimental results about heavy quarkonia at finite temperature.

In this contribution we present a new approach~\cite{Caselle:2019tiv}, based on conformal perturbation theory (CPT), to study the confining potential when $T$ is close to the deconfinement temperature. Our analysis holds for second-order deconfinement transitions: as such, it cannot be used for pure-glue $\SU(3)$ Yang-Mills theory (which has a first-order deconfinement phase transition), but it could be used to describe the physics near the critical endpoint in the QCD phase diagram at finite temperature and at finite chemical potential $\mu$~\cite{Lacey:2014wqa}. In particular, this critical point is expected to be in the same universality class as the one associated with the thermal deconfinement transition of $\SU(2)$ Yang-Mills theory that we discuss here, hence most of our results could be relevant for it. As is well known, the $\SU(2)$ Yang-Mills theory has a second order deconfinement transition in the same universality class of the 3d Ising model~\cite{Svetitsky:1982gs}; for this theory, it is easy to obtain high-precision lattice results for the interquark potential and carry out robust tests of our analytical predictions based on CPT.

\section{Conformal perturbation theory}
\label{sec:CPT}

Conformal perturbation theory allows one to construct a perturbative expansion for the short-distance behavior of various correlation functions in the vicinity of a conformally invariant critical point, based only on the properties of the model at its critical point~\cite{Guida:1995kc,Guida:1996nm,Caselle:2001zd,Caselle:1999mg,Amoretti:2017aze}. Some of this information (e.g. the structure constants appearing in operator-product expansions and the critical indices) is universal and can be obtained, for instance, using the bootstrap approach. On the other hand, quantities such as amplitudes of one-point functions are non-universal, and can be  obtained using techniques like strong- or weak coupling expansions or lattice simulations. Thanks to the recent progress in the bootstrap approach, the universal quantities are presently known with high precision for several universality classes. This enables one to obtain very precise predictions for off-critical correlators, in particular for the Ising model in three dimensions~\cite{Caselle:2015csa,Caselle:2016mww}. In this case, the predictive power of CPT is enhanced by the fact that the operator content of the model is very simple, with well-spaced conformal weights. In particular, there exists a large gap between the first three terms in the expansion and the remaining ones, so that a truncated CPT expansion including only these three terms yields excellent agreement both with lattice estimates of the correlators and with experimental results~\cite{Caselle:2016mww}. Moreover, one finds that the second and third term have similar weights and opposite signs, and that both of them have to be included in the analysis. Since the deconfinement transition of $\SU(2)$ Yang-Mills theory in $3+1$ spacetime dimensions is in the same universality class of the tridimensional Ising model, the same methods are expected to be relevant also for this theory. In particular, we study the behavior of the two-point correlation function of Polyakov loops at short distances. As the Polyakov loop is the order parameter of the deconfinement transition, this correlation function is mapped to the spin-spin correlator $\langle \sigma \sigma \rangle$ of the Ising model. The CPT expansion of this correlator is well known (see for instance \cite{Caselle:2016mww}):
\begin{equation}\label{corr1}
\langle \sigma (r) \sigma (0) \rangle _t = C_{\sigma\sigma}^{1} (0,r) + 
C_{\sigma\sigma}^{\epsilon} (0,r) A^{\pm} |t|^{\frac{\Delta_{\epsilon}}{\Delta_{t}}} +
t  \partial_{t} C_{\sigma\sigma}^{1} (0,r) + ... ,
\end{equation}
where  $\Delta_{\sigma} = 0.5181489 (10) $ and $\Delta_{\epsilon} = 1.412625 (10) $ are the conformal weights of the spin and energy ($\sigma$ and  $\epsilon$, respectively) operators
\cite{Komargodski:2016auf,Kos:2016ysd}, and $\Delta_{t}=3-\Delta_{\epsilon}$. $A^{\pm}$ denotes the non-universal critical amplitude of the energy operator above ($+$) and below ($-$) the critical point,
\begin{equation}
\langle \epsilon \rangle_t = A^{\pm} |t|^{\frac{\Delta_{\epsilon}}{\Delta_{t}}},
 \end{equation}
 whereas
\begin{equation}
C_{\sigma\sigma}^{1}(0,r) = \frac{1}{r^{2 \Delta_ \sigma}},~~~~~  C_{\sigma\sigma}^{\epsilon}(0,r) = C_{\sigma\sigma}^{\epsilon} r^{\Delta _\epsilon -2 \Delta _\sigma}
\end{equation} 
denote the Wilson coefficients evaluated at the critical point (having fixed the conventional normalization $C_{\sigma\sigma}^{1}=1$), and $\partial_{t} C_{\sigma\sigma}^{1} (0,r)$ is the derivative of the Wilson coefficient with respect to the perturbing parameter $t$, evaluated at the critical point. This derivative can be evaluated explicitly, leading to the following \emph{universal} result:
\begin{equation}
\partial _t C^1 _{\sigma  \sigma } (0,r) \equiv   62.5336... r^{\Delta _t - 2 \Delta _\sigma} C^{\epsilon}_{\sigma\sigma}.
\end{equation}
The latter expression can be rewritten in a more elegant way in terms of the scaling variable  $s = t \cdot r^{\Delta_{t}}$, which is the actual expansion parameter of CPT:
\begin{equation}\label{corr1s}
r^{2\Delta_{\sigma}} \langle  \sigma (r) \sigma (0)\rangle _t = 1 + 
C_{\sigma\sigma}^{\epsilon} A^{\pm} |s|^{\frac{\Delta_{\epsilon}}{\Delta_{t}}} + 62.5336...
C_{\sigma\sigma}^{\epsilon}  s .
\end{equation}
In order to use this result to describe the short-distance behavior of Polyakov-loop correlators, one must still fix a set of non-universal quantities:
\begin{enumerate}
\item The normalization of the Polyakov loop, i.e. the proportionality factor relating the spin operator $\sigma$ appearing in the equations above and the Polyakov loop $P$ evaluated in our simulations. This proportionality factor can be obtained from the two-point correlation function at the critical point.
\item The conversion factor relating the lattice and continuum versions of the scaling parameter $s$. The simplest way to perform this conversion is to fix $R=r$, (i.e. to identify the interquark distance measured in units of the lattice spacing on the lattice $R$ with the distance $r$ in continuum appearing in the equations above) and include all non-universal constants in the conversion between the reduced temperature $t$ in the continuum and the perturbing parameter of the $\SU(2)$ gauge theory, i.e. the difference $\beta-\beta_{\mbox{\tiny{c}}}(N_t)$, where $\beta=4/g^2$ is the Wilson parameter, inversely proportional to the bare gauge coupling $g$, and $\beta_{\mbox{\tiny{c}}}(N_t)$ denotes the value it takes at the critical temperature on a lattice with $N_t$ sites in the Euclidean-time direction. In the following we rewrite, as usual, this difference in terms of the dimensionless ratio $T/\Tc$, using the scale-setting relation reported in ref.~\cite{Caselle:2015tza}.
To define this relation we use the fact that the last term in eq.~(\ref{corr1s}) is universal:  we fit our results for the correlator as a function of $r$, and use the result to fix the relation between $t$ and $T/\Tc$. 
\item Finally, we can compute $A^{\pm}$ by using the second term of the expansion. The numerical value of these amplitudes is an interesting output of our analysis, in addition to the functional form of the correlator. As a non-trivial test of the whole procedure, we also check that the $A^+/A^-$ ratio is universal, as expected.
\end{enumerate}

\section{Numerical results for the $3+1$ dimensional $\SU(2)$ lattice gauge theory}
\label{sec:su2results}

We tested the predictions discussed in the previous section by comparing them with lattice results for Polyakov-loop correlators in the vicinity of the finite-temperature deconfinement transition of $\SU(2)$ Yang-Mills theory in $3+1$ spacetime dimensions. We carried out this study for temperatures below and above the critical temperature $\Tc$. The theory is regularized on a finite lattice of spacing $a$ and spacetime volume $\mathcal{V}=a^4 N_t \times N_s^3$ with periodic boundary conditions in all directions. The action is taken to be~\cite{Wilson:1974sk}
\begin{equation}
\SW = -\frac{2}{g^2} \sum_{x} \sum_{0 \le \mu < \nu \le 3} \Tr U_{\mu\nu} (x),
\end{equation}
where $U_{\mu\nu} (x)$ is the plaquette associated with the site $x$ in the directions $\mu$ and $\nu$, and $g$ is the bare coupling. As mentioned above, we also define the Wilson parameter $\beta=4/g^2$. The temperature $T$ is given by the inverse of the extent $aN_t$ of the shortest compactified side of the lattice, while the sizes of the three ``spatial'' directions are much larger ($N_s \gg N_t$). In order to vary the temperature continuously, one can tune the lattice spacing $a$ by tuning the Wilson parameter: for details on the scale-setting procedure, we refer to ref.~\cite{Caselle:2015tza}. We also use the fact that, for this theory, the deconfinement temperature and the zero-temperature string tension $\sigma$ are related to each other by $\Tc/\sqrt{\sigma}=0.7091(36)$~\cite{Lucini:2003zr}. Table~\ref{tab:lattice_setup} reports the setup of our Monte~Carlo calculations.

\begin{table}
\centering
\begin{tabular}{|c|c|c|c|c|}
\hline
$\beta$ & $ N_t \times N_s^3 $ & $T/\Tc$ & $\nconf$ \\
\hline
\hline
$2.48479$  & $8\times 80^3$ & $0.90$ & $8\times10^4$ \\
$2.50311$  & $8\times 80^3$ & $0.96$ & $8\times10^4$ \\
$2.50598$  & $8\times 80^3$ & $0.97$ & $8\times10^4$ \\
$2.51165$  & $8\times 80^3$ & $1$    & $8\times10^4$ \\
$2.52295$  & $8\times 80^3$ & $1.02$ & $8\times10^4$ \\
$2.52567$  & $8\times 80^3$ & $1.05$ & $8\times10^4$ \\
$2.54189$  & $8\times 80^3$ & $1.10$ & $8\times10^4$ \\
\hline
\hline
$2.55$    & $10\times 80^3$ & $0.90$ & $10^5$ \\
$2.569$   & $10\times 80^3$ & $0.96$ & $10^5$ \\
$2.572$   & $10\times 80^3$ & $0.97$ & $10^5$ \\
$2.58101$ & $10\times 80^3$ & $1$    & $8\times10^4$   \\
$2.58984$ & $10\times 80^3$ & $1.02$ & $1.6\times10^5$ \\
$2.59271$ & $10\times 80^3$ & $1.05$ & $1.6\times10^5$ \\
$2.61$    & $10\times 80^3$ & $1.10$ & $1.6\times10^5$ \\
\hline
\hline
$2.60573$    & $12\times 96^3$ & $0.90$ & $8\times10^4$   \\
$2.626$      & $12\times 96^3$ & $0.96$ & $8\times10^4$   \\
$2.62923$    & $12\times 96^3$ & $0.97$ & $8\times10^4$   \\
$2.63896$    & $12\times 96^3$ & $1$    & $8\times10^4$   \\
$2.64558$    & $12\times 96^3$ & $1.02$ & $1.6\times10^5$ \\
$2.65541$    & $12\times 96^3$ & $1.05$ & $1.6\times10^5$ \\
$2.67085$    & $12\times 96^3$ & $1.10$ & $1.6\times10^5$ \\
\hline
\end{tabular}
\caption{\label{tab:lattice_setup} Setup of the lattice calculations performed for the $\SU(2)$ Yang-Mills theory: the first column shows the values of the Wilson parameter $\beta=4/g^2$, the second displays the temporal and spatial sizes, in the third column we report the corresponding temperature in units of $\Tc$, and finally in the last column the statistics of  measurements of Polyakov loop correlators are given.}
\end{table}
The Polyakov loop at a spatial point $\vec{x}$ is defined as
\begin{equation}
\label{Polyakov_loop}
 P\left(\vec{x}\right) = \frac{1}{2} \Tr \prod_{0 \le t < N_t} U_0 (\vec{x},t a) \; .
\end{equation}
The observable we are interested in is the two-point correlation function of Polyakov loops at a spatial separation $R$ (along one of the main spatial lattice axes $\hat{k}$), i.e.
\begin{equation}
\label{def_G}
 G(R) = \left\langle \sum_{\vec{x}} \ P\left(\vec{x}\right) P\left(\vec{x}+R \hat{k}\right) \right\rangle .
\end{equation}
Note that the Polyakov loop defined in eq.~(\ref{Polyakov_loop}) (and, as a consequence, also its correlators) is a \emph{bare} quantity, that vanishes in the continuum limit $a \to 0$ and requires renormalization: see ref.~\cite{Mykkanen:2012ri} and references therein for a thorough discussion.

\subsection{Comparison with CPT predictions}
\label{sec:Comparison_CPT}

\begin{figure}
\centerline{\includegraphics{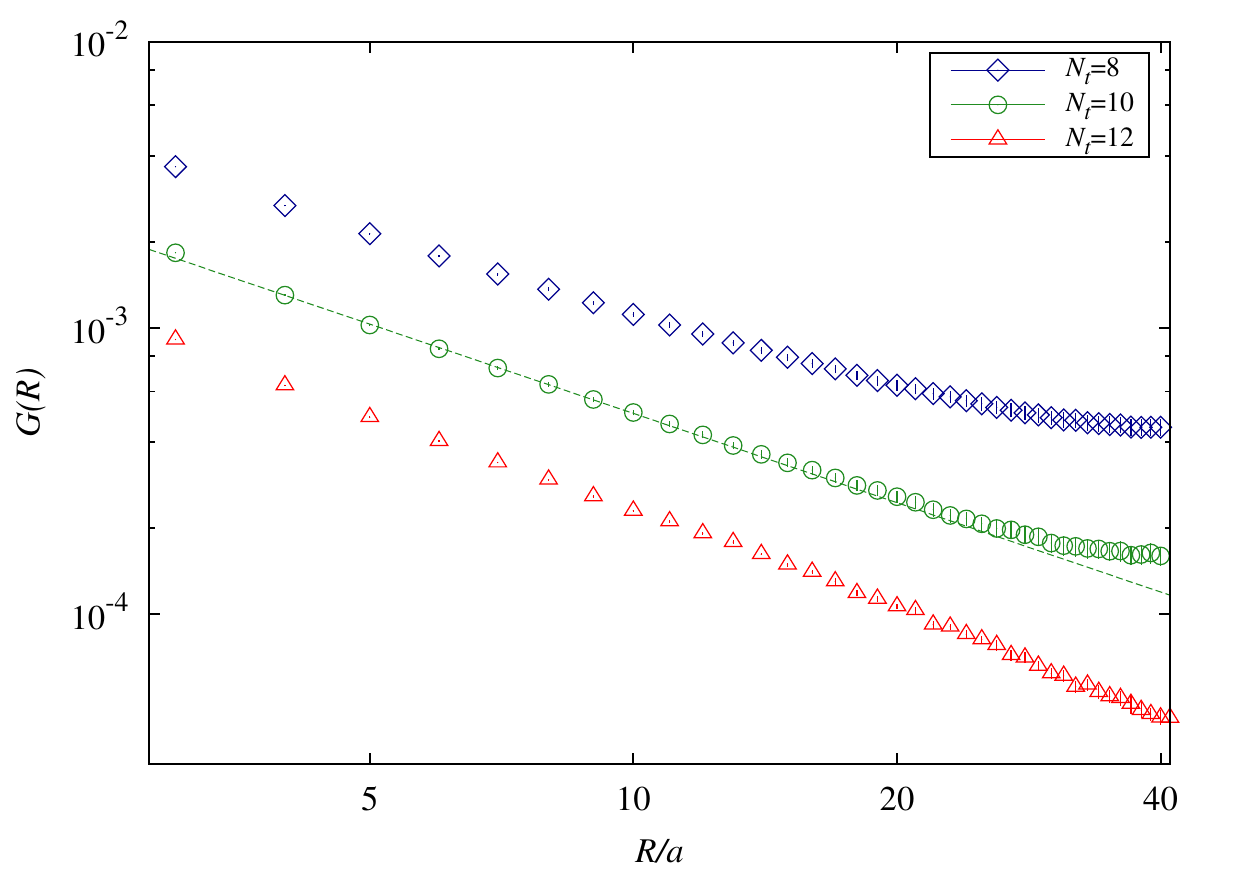}}
\caption{\label{Fig_T_c} Results for the Polyakov loop correlator as a function of the interquark distance $R$ in units of $a$ at the deconfinement point ($T=\Tc$). Note that both axes are displayed in logarithmic scale.}
\end{figure}

We analyzed our Monte~Carlo results and compared them with CPT predictions following this procedure:
\begin{enumerate}
\item First, we studied the behavior of the correlator at the critical point so as to extract the normalization constant for the Polyakov loops.
\item Using this constant as input we fitted the numerical value of the correlator with eq.~\ref{corr1s} as a function of $R$, keeping the coefficients of the second and third term in the expansion as free parameters.
\item Finally, we used our best estimates for these coefficients to fix the remaining non-universal quantities and to investigate their dependence on $T$.
\end{enumerate}
The values of the correlator at the critical point are reported in fig.~\ref{Fig_T_c}: the figure shows that the data agree with the expected power-law behavior (a straight line in this plot, where both axes are displayed in a logarithmic scale) for values of $R$ in the range $3a<R<20a$, while  for larger values of $R$ the data are affected by effects due to the finite size of the lattice.

We extracted the normalization of the Polyakov loop fitting  the data in the range $3a<R<20a$ according to the law
\begin{equation}
\label{fit1}
 G(R)=\frac{C_P^2}{(R/a)^{2\Delta_{\sigma}}}
\end{equation}
using $C_P$ as the only free parameter we found $C_P^2=0.01070(1)$ , $C_P^2=0.00547(2)$  and $C_P^2=0.00253(1)$ for the data from simulations performed respectively at $N_t=8$, $N_t=10$, and $N_t=12$, that are shown in the figure. Further details on the fitting procedure will be reported in a forthcoming publication~\cite{Caselle:2019tiv}.

\begin{figure}
\centerline{\includegraphics{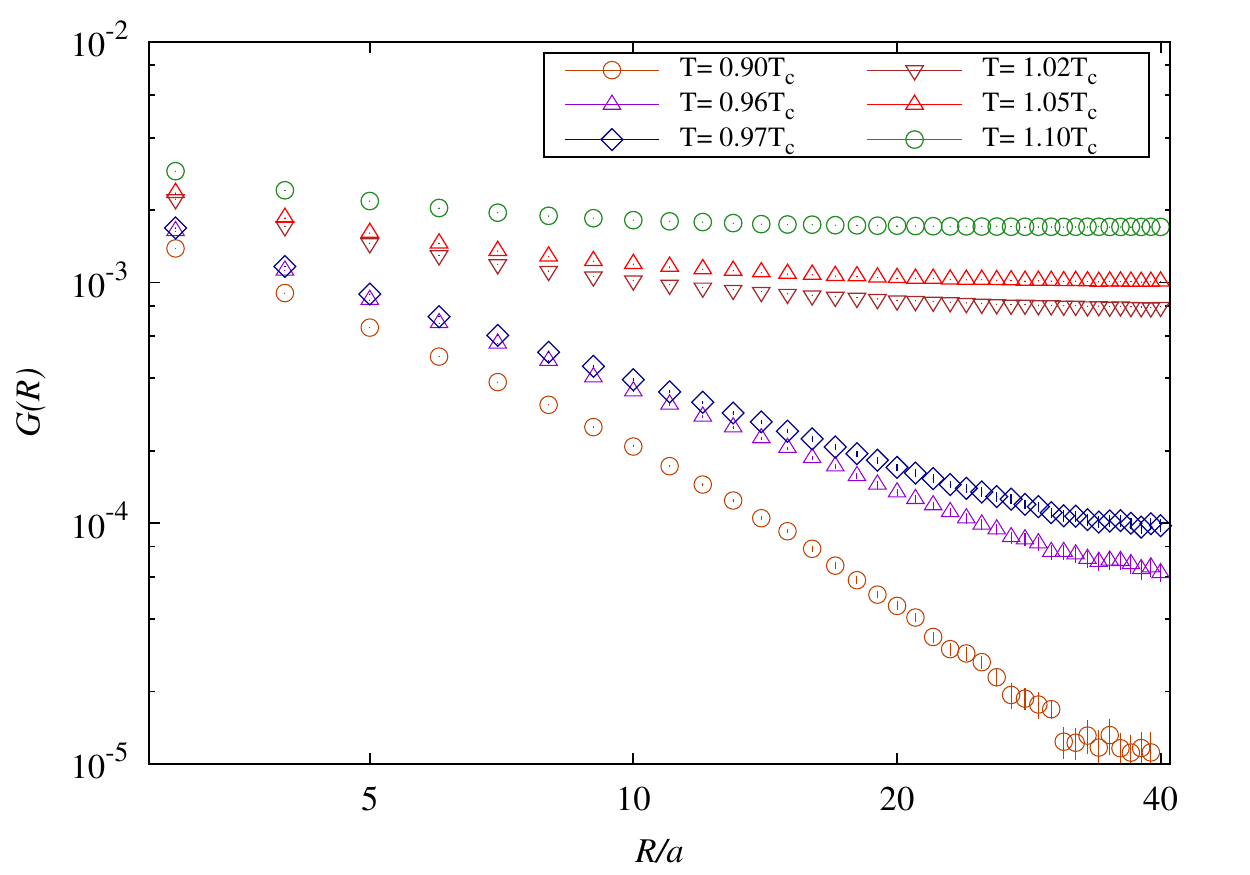}}
\caption{\label{Fig2} Results for the Polyakov loop correlator as a function of the interquark distance $R$ (in units of the lattice spacing $a$) at $T \neq \Tc$, for different values of the temperature.}
\end{figure}

We then fitted the Polyakov loop correlators $G(R)$ at $T \neq \Tc$ using the functional form
\begin{equation}
\label{fit2}
G(R) = \frac{C^2_P}{(R/a)^{2\Delta_{\sigma}}} \left[ 1 + c \cdot (R/a)^{\Delta_{\epsilon}} + b \cdot (R/a)^{\Delta_{t}} \right]
\end{equation}
in which the coefficients $\Delta_{\sigma}, \Delta_{\epsilon}$ and $\Delta_{t}$ are those discussed in section~\ref{sec:CPT} and $b$ and $c$ are the fit parameters. As an example, in table~\ref{results} we report the results of this analysis for the $N_t=10$ case. The typical shape of these correlators and their deviation from a pure power law for $T \neq \Tc$ 
can be appreciated from fig.~\ref{Fig2}, where we report them for $N_t=10$, using a logarithmic scale for the axes. Also in this case we fitted the data for different ranges of distances, up to $R/a\in[4,20]$, finding values of the reduced $\chi^2$ of order one in all cases.

\begin{table}
\begin{center}
\begin{tabular}{|c|c|c|c|c|c|}
\hline
$\beta$ & $T/\Tc$ & $c$& $b$ \\
\hline
$2.55$    & $0.90$  & $-0.169(1)$ & $0.099(1)$    \\
$2.569$   & $0.96$  & $-0.067(2)$ & $0.037(1)$    \\
$2.572$   & $0.97$  & $-0.048(3)$ & $0.026(2)$    \\
$2.58984$ & $1.02$  & $0.067(2)$  & $-0.019(1)$   \\
$2.59271$ & $1.05$  & $0.091(2)$  & $-0.0256(15)$ \\
$2.61$    & $1.10$  & $0.221(3)$  & $-0.081(3)$   \\
\hline
\end{tabular}
\caption{\label{results} Results of the fits of the correlator $G(R)$ for different values of the Wilson parameter $\beta$ (first column), corresponding to the temperatures reported in the second column: the last two columns display the fitted values for the two parameters $c$ and $b$.}
\end{center}
\end{table}

The best-fit estimates of $b$ can be used to define the value of the perturbing parameter $t$ corresponding to each value of $\beta$ (and thus of $T/\Tc$) via
\begin{equation}
b= -62.5336 \, t \, C^{\epsilon}_{\sigma\sigma}.
\end{equation}
Inserting the value for the structure constant $C^{\epsilon}_{\sigma \sigma} = 1.0518537(41)$ known from the literature~\cite{Komargodski:2016auf,Kos:2016ysd} we end up with the results for $t$ reported in table~\ref{tab_A} for $N_t=10$, in table~\ref{tab_B} for $N_t=8$, and finally in table~\ref{tab_C} for $N_t=12$.

As expected, $t<0$ in the deconfined phase (where the $\mathbb{Z}_2$ center symmetry of the Yang-Mills theory and, correspondingly, the $\mathbb{Z}_2$ symmetry of the Ising model get broken) while $t>0$ in the confining ($\mathbb{Z}_2$-symmetric) phase. The values of $t$ turn out to be slightly larger than, but of the same order of magnitude as, those studied in ref.~\cite{Caselle:2016mww}. This gives confidence that also in this case the truncated CPT approximation should give quantitatively accurate results.

From $t$, one can extract the non-universal amplitudes $A^\pm$ using
\begin{equation}
A^\pm=\frac{c}{C^{\epsilon}_{\sigma\sigma}}t^{-\frac{\Delta_{\epsilon} }{\Delta_{t} }}\,.
\end{equation}
The determination of these two constants allows for some non-trivial tests of the whole procedure: in particular, $A^+$ must have a constant negative value for all the values of $\beta$ in the confining phase, while $A^-$ must have the same value in all the simulations performed in the deconfined phase. Moreover, even though neither amplitude is universal, their ratio \emph{is}.

Our results for $A^\pm$ are reported in the last column of tables~\ref{tab_A}, \ref{tab_B}, and~\ref{tab_C}: they show that in the confining phase the values are stable, compatible with each other within the statistical uncertainties, and show no dependence on $N_t$. By contrast, in the deconfined phase they are noisier, but the differences between the various values are always smaller than $10\%$, which (given the non-trivial numerical analysis from which they are extracted and the fact that these values are derived from correlators in the broken-symmetry phase) is nevertheless a reasonably good result. Combining all results, we end up with an estimate $A^+=-52(4)$ in the confining phase and $A^-=84(8)$ in the deconfined phase. These values correspond to a ratio of $-A^+/A^-=0.62(10)$, which turns out to be perfectly compatible with the estimate $-A^+/A^-=0.536(2)$ previously reported in ref.~\cite{Hasenbusch:2010ua}.

\begin{table}
\begin{center}
\begin{tabular}{|c|c|c|c|}
\hline
$\beta$ & $T/\Tc$ & $t$ &$A^\pm$ \\
\hline
$2.55$    & $0.90$ & $0.001505(15)$  & $-52.6(6)$ \\
$2.569$   & $0.96$ & $0.000563(15)$  & $-50(2)$   \\
$2.572$   & $0.97$ & $0.000395(30)$  & $-49(5)$   \\
$2.58984$ & $1.02$ & $-0.000284(18)$ & $91(6)$    \\
$2.59271$ & $1.05$ & $-0.000389(23)$ & $94(5)$    \\
$2.61$    & $1.10$ & $-0.001231(45)$ & $82(3)$    \\
\hline
\end{tabular}
\caption{\label{tab_A} Results obtained from simulations on lattices with $N_t=10$ for the perturbing parameter $t$ and for the amplitude $A^+$ in the confining phase (first three rows) and for $A^-$ in the deconfining phase (last three rows), for the values of the Wilson parameter $\beta$ and at the temperatures $T$ in units of the deconfinement temperature $\Tc$ reported in the first and second column, respectively.}
\end{center}
\end{table}

\begin{table}
\begin{center}
\begin{tabular}{|c|c|c|c|}
\hline
$\beta$ & $T/\Tc$ & $t$ & $A^\pm$ \\
\hline
$2.48479$ & $0.90$ & $0.001429(15)$  & $-53.0(6)$ \\
$2.50311$ & $0.96$ & $0.000429(15)$  & $-54(2)$   \\
$2.50598$ & $0.97$ & $0.000350(14)$  & $-52(2)$   \\
$2.52295$ & $1.02$ & $-0.000678(2)$  & $83.0(2)$  \\
$2.52567$ & $1.05$ & $-0.000841(2)$  & $85.6(2)$  \\
$2.54189$ & $1.10$ & $-0.001802(15)$ & $84.7(7)$  \\
\hline
\end{tabular}
\caption{\label{tab_B} Same as in table~\ref{tab_A}, but from simulations on lattices with $N_t=8$.} 
\end{center}
\end{table}

\begin{table}
\begin{center}
\begin{tabular}{|c|c|c|c|}
\hline
$\beta$ & $T/\Tc$ & $t$ & $A^\pm$ \\
\hline
$2.60573$ & $0.90$ & $0.001019(30)$  & $-53.3(1.7)$ \\
$2.626$   & $0.96$ & $0.000228(12)$  & $-51(3)$     \\
$2.62923$ & $0.97$ & $0.000149(2)$   & $-51(1)$     \\
$2.64558$ & $1.02$ & $-0.000331(12)$ & $74(3)$      \\
$2.65541$ & $1.05$ & $-0.000617(5)$  & $84.7(9)$    \\
$2.67085$ & $1.10$ & $-0.00158(11)$  & $74(5)$      \\
\hline
\end{tabular}
\caption{\label{tab_C} Same as in table~\ref{tab_A}, but from simulations on lattices with $N_t=12$.}
\end{center}
\end{table}

\subsection{Concluding remarks}

In this contribution, we presented a novel analytical method, based on conformal perturbation theory, to study the interquark potential in non-Abelian gauge theories at finite temperature, and we tested it against a new set of high-precision lattice results for two-point correlation functions of Polyakov loops in $\SU(2)$ Yang-Mills theory. As we remarked, this gauge theory has a second-order deconfinement transition, in the same universality class relevant for the critical endpoint of the QCD phase diagram~\cite{Lacey:2014wqa}, i.e. the one of the Ising model in three dimensions.

Our fits of the lattice results to the truncated CPT formula in eq.~(\ref{fit2}) always result in values of the reduced $\chi^2$ of order one, i.e. indicate that the lattice results are described very well by CPT. These results confirm the findings of an analogous investigation that was carried out for the correlators evaluated numerically in the Ising model~\cite{Caselle:2016mww}.

We remark that the CPT prediction works well even at the lowest temperature that we studied, $T/\Tc=0.9$, which is quite far from the critical point. We could not probe lower temperatures  because for $T<0.9\Tc$ the correlation length in units of the lattice spacing becomes too short, even for the finest lattices (corresponding to $N_t=12$) that we studied. However, it is possible that, increasing $N_t$, one could find good agreement with CPT at even lower temperatures.

We argued that the fact that truncated CPT predictions are so accurate is not accidental: rather, it can be interpreted in terms of the large gap between the first terms of the CPT expansion, that are explicitly included in eq.~(\ref{fit2}), and higher-order terms, that we are neglecting.

This peculiar feature of the Ising universality class makes the CPT approach very useful to study the dynamics of systems in this universality class, including, in particular, the critical endpoint of the QCD phase diagram at finite temperature and finite baryonic chemical potential.

\end{document}